\documentstyle[a4,12pt,amssymb,float,epsfig,psfig,times]{article}
\newcommand{\be}{\begin{equation}}
\newcommand{\ee}{\end{equation}}
\newcommand{\bea}{\begin{eqnarray}}
\newcommand{\eea}{\end{eqnarray}}

\def\Tu{\left< T^f_{uu} \right>}

\def\p1{\pi_1}

\def\d{\delta}

\renewcommand{\thefootnote}{\fnsymbol{footnote}}

\begin{document}

\begin{titlepage}

\vspace*{\stretch{0}}

\begin{center}
{\Large\bf A Planck-like problem for quantum charged black holes}
\\[0.5cm]
A. Fabbri$^{\rm a}$\footnote{\tt fabbria@bo.infn.it},
D. J. Navarro$^{\rm b}$\footnote{\tt dnavarro@ific.uv.es} and
J. Navarro-Salas$^{\rm b}$\footnote{\tt jnavarro@ific.uv.es}
\\[0.5cm]
{\footnotesize
a) Dipartimento di Fisica dell'Universit\`a di Bologna and INFN
sezione di Bologna,\\
Via Irnerio 46, 40126 Bologna, Italy.
\\[0.5cm]
b) Departamento de F\'{\i}sica Te\'orica and IFIC, Centro Mixto
Universidad
de Valencia-CSIC.\\
Facultad de F\'{\i}sica, Universidad de Valencia, Burjassot-46100,
Valencia,
Spain.}
\end{center}
\bigskip
\begin{center}
{\it Awarded Fifth Prize in the Gravity Research Foundation Essay 
Competition for 2001}
\end{center}
\bigskip
\begin{abstract}
Motivated by the parallelism existing between
the puzzles of classical physics at the beginning of the XXth
century and the current paradoxes in the search of a  quantum
theory of gravity, we give, in analogy with Planck's black body
radiation problem, a solution for the exact Hawking flux of
evaporating Reissner-Nordstr\"om black holes. Our results show
that when back-reaction effects are fully taken into account the
standard picture of black hole evaporation is significantly
altered, thus implying a possible resolution of the information
loss problem.
\end{abstract}
\end{titlepage}
\newpage
\renewcommand{\thefootnote}{\arabic{footnote}}
\setcounter{footnote}{0}

The remarkable discovery that black holes emit thermal radiation
 \cite{h1} has raised serious doubts on the unitarity of a quantum
 theory of gravity. Hawking argued \cite{h2} that
the semiclassical approximation should be valid until the Planck
mass is reached. This, in turn, implies that the black hole should
shrink slowly during the evaporation. At the Planck mass there is
not enough energy inside the black hole to radiate out the
information of the collapsed matter, thus implying a loss of
quantum coherence. However, it has been stressed \cite{th} that
gravitational back-reaction effects could change the standard
picture of the evaporation process. It is clear that the
back-reaction must be very important, at least at late times, in
order to prevent the total emitted energy to diverge. In contrast,
at early times one could expect these effects to be negligible and
the radiation can be calculated using the classical space-time
geometry.\\

A natural scenario where one can exactly evaluate the emitted
radiation at late times is in the scattering of extremal
Reissner-Nordstr\"om (RN) black holes by massless neutral
particles. If the incoming matter has a long-wavelength, and
preserves spherical symmetry, the resulting near-extremal RN black
hole can be described, by an infalling observer crossing the
horizon, by means of the ingoing Vaidya-type metric \cite{fnn1}
\be
\label{inmetric} ds^2 = -\frac{2l}{r_0} \left(
\frac{2x^2}{l^2q^3}-lm(v) \right) dv^2 + 2 dv dx + (r_0^2 + 4lx)
d{\Omega}^2 \, , \ee where $l^2=G$ is Newton's constant and
$r_0=lq$ is the extremal radius. The function $m(v)$ represents
the deviation of the mass from extremality and it verifies the
evolution law
\be
\label{mvequation}
\partial_v m(v) = -\frac{\hbar}{24\pi lq} m(v) + (\partial_v f)^2 \, ,
\ee $f$ being the null matter field related to the 2d stress
tensor $T_{vv}^f$, and the corresponding 4d one $T_{vv}^{(4)}$, by
$(\partial_v f)^2 \equiv T^f_{vv} = 4\pi r^2 T_{vv}^{(4)}$.
Technically, this is the solution coming from the effective action
$S_{eff}$ which is obtained by integrating out the field $f$. Due
to spherical symmetry and considering the near-horizon region,
$S_{eff}$ corresponds to the classical action plus the
Polyakov-Liouville one \cite{p}.\\

In the absence of incoming matter ($m(v)=0$), and in the very
near-horizon limit $r_0^2>>lx$, the metric (\ref{inmetric})
recovers the Robinson-Bertotti anti-de Sitter geometry \cite{rb}.
In the dynamical situation the metric (\ref{inmetric}) implies the
existence of a negative incoming quantum flux crossing the horizon
that goes down exponentially for $v \rightarrow +\infty$. A
``complementary'' description can also be given  from the point of
view of an asymptotic observer, for whom there is no incoming
radiation but there exists an outgoing (Hawking) evaporation flux.
The geometry for the outside observer can be then described, at
late times, by an outgoing Vaidya-type metric \cite{fnn2}
\be
\label{outmetric}
ds^2 \sim -\frac{2l}{r_0} \left( \frac{2x^2}{l^2q^3}-lm(u) \right) dv^2
-
2 du dx + (r_0^2 + 4lx) d{\Omega}^2 \, ,
\ee
where $m(u)$ satisfies the equation
\be
\label{muequation}
\partial_u m(u) = -\frac{\hbar}{24\pi lq^3} m(u) \, ,
\ee
which can be integrated easily
\be
m(u) = m_0 e^{-\frac{N\hbar}{24\pi lq^3} u} \, . \ee Requiring
that both descriptions match at the end-point $u=+\infty,
v=+\infty$ (where the extremal configuration is recovered) we can
determine the integration constant $m_0$
\be
m_0 = m(v_f)e^{\frac{N\hbar}{24\pi lq^3} v_f} \, ,\ee where $v_f$
is the value of the advanced time $v$ at which the classical
incoming matter is turned off and $m(v_f)$ is calculated using
(\ref{mvequation}) with the condition $m(v=-\infty)=0$. For finite
values of $u$ and $v$ the descriptions of the two observers differ
and this, in some sense, is in agreement with the principle of
complementarity \cite{th,stu}. Moreover, this expression gives
immediately the exact asymptotic behaviour of the Hawking flux at
late times $u \rightarrow +\infty$
\be
\label{lateflux} \left< T^f_{uu}(u) \right> \sim
\frac{\hbar}{24\pi lq^3} m(v_f) e^{-\frac{N\hbar}{24\pi lq^3}
(u-v_f)} \, . \ee Note that two different expansions in $\hbar$
are implicit in (\ref{lateflux}). One is associated to the
exponential, fulfilling Stefan's law, and the other inside
$m(v_f)$. The second is physically very interesting because it
captures details of the incoming matter.\\

The issue is now to work out the Hawking radiation flux
for finite $u$, but to this end we can no longer use the one-loop
effective action that controls the physics near the horizon.
Therefore we face directly with the problem that the effective
theory describing the whole asymptotic region is unknown. However,
the physical intuition suggests the following set of conditions
for the Hawking flux $\Tu$ (with back-reaction effects included):
\begin{enumerate}
\item At early times, $u \rightarrow -\infty$, it must coincide with
the Hawking flux calculated ignoring the back-reaction.
\item At late times, $u \rightarrow +\infty$, it should behave
as (\ref{lateflux}).
\item At leading order in $\hbar$, it must also agree with
the Hawking flux computed neglecting the back-reaction.
\item It has to be compatible with energy conservation.
\end{enumerate}

Therefore we deal with a problem similar to that considered by
Planck, one century ago, for the black body radiation and  the
conditions 1 and 2 mimic, respectively, the Rayleigh-Jeans and
Wien laws for low and high frequencies of black body emission.
They represent small and large back-reaction effects or, in the
black body analogy, classical and pure quantum behaviours.
Such as in this analogy, what we need now is an interpolating
function matching the asymptotic behaviours at early and late
times. Note that in our problem we have two additional
requirements: at leading order in $\hbar$ the flux should agree
with that calculated in the fixed classical background (condition
3) and it has to be compatible with energy conservation (condition
4).\\

We shall now provide a solution to this problem if the incoming
matter is given by a finite set of spherical null shells with
energies $m_1, \ldots m_N$ falling into the black hole at the
advanced times $v_1, \ldots v_N$, respectively. Note that in the
limit $N \rightarrow \infty$, while keeping $N(v_N - v_{1})$
finite, we can imitate a continuous distribution of matter. In
this situation the classical stress tensor is given by $T^f_{vv} =
\sum_{i=1}^N m_i \d (v-v_i)$, and it can be written as
\be
T^f_{vv} = \sum_{i=1}^N m a_i \d (v-v_i) \, ,
\ee
where $a_i = \frac{m_i}{m}$, $m = \sum_{i=1}^N m_i$.
The classical solution can be determined by matching static solutions at
$v_1, \ldots v_N$. In this way one can write the relation between
the initial $u_{in}$ and final $u$ outgoing Eddington-Finkelstein
coordinates \bea \frac{du}{du_{in}} &=&
\frac{(r(v_N,u_{N-1})-r_+^{(N-1)})(r(v_N,u_{N-1})-r_-^{(N-1)})}
{(r(v_N,u)-r_+^{(N)})(r(v_N, u)-r_-^{(N)})}
\frac{(r(v_{N-1},u_{N-2})-r_+^{(N-2)})}{(r(v_{N-1},u_{N-1})
-r_+^{(N-1)})}
 \nonumber\\ & &
\frac{(r(v_{N-1},u_{N-2})-r_-^{(N-2)})}
{(r(v_{N-1},u_{N-1})-r_-^{(N-1)})}......
\frac{(r(v_1,u_{in})-q)^2}{(r(v_1,u_1)-r_+^{(1)})(r(v_1,u_1)-r_-^{(1)})}
\, , \eea where $r(v,u_i)$ is the radial function after the i-th
null shell and $r_{\pm}^{(i)}$ are the outer and inner horizons of
the corresponding static RN black hole.\\

The Hawking flux without back-reaction $\Tu_{NBR}$ is given by the
Schwarzian
derivative \cite{bd}
\be
\left< T^f_{uu}(u) \right>_{NBR} = -\frac{\hbar}{24\pi} \{ u_{in},
u\} \, , \ee and it can be regarded as a function of $u$, $m$,
$v_i$ and $a_i$. Let us now consider the differential equation
\be
\frac{dm(u)}{du} = -\Tu_{NBR} (u, m(u), v_i, a_i) \, , \ee with
the initial condition $m(u=-\infty) = \sum m_i = m$, and
substitute the constant $m$ for $m(u)$ in the Hawking flux
calculated on the fixed classical background $\Tu_{NBR}$. In this
way we get a Hawking flux $\Tu$ verifying the above four
conditions.\\

For the asymptotic observer the effective metric is given by an
outgoing Vaidya metric with mass $m(u)$. Energy conservation
requires that
\be
\frac{dm(u)}{du} = - \Tu \, ,\ee and therefore condition 4 is
satisfied by construction. At early times $u \rightarrow -\infty$
and also at leading order in $\hbar$ for all $u$ (no
back-reaction), we have $m(u)=m$ and therefore $\Tu=\Tu_{NBR}$. At
late times $u \rightarrow +\infty$ the solution $m(u)$ behaves as
\be
m(u) \sim \sum_i m_i e^{-\frac{\hbar}{24\pi q^3}(u-v_i)} \, , \ee
in agreement with condition 2. It is well known that, without
back-reaction effects, the Hawking flux $\Tu_{NBR}$ approaches the
constant thermal value very rapidly, as soon as the exponential
late time form of the redshift factor is reached
\be
\label{thermal} \frac{du}{du_{in}} \sim e^{2\pi T_H u} \, , \ee
where $T_H$ is the Hawking temperature. Once we consider
back-reaction effects, due to the interaction between infalling
matter and outgoing radiation the relation between the coordinates
$u_{in}$ and $u$ is given, in terms of our exact evaporation flux,
by the differential equation
\be
\left< T^f_{uu}(u) \right> = -\frac{\hbar}{24\pi} \{u_{in}, u\} \,
. \ee Because $\left< T^f_{uu}(u) \right>\to 0$ at $u\to
\pm\infty$,   thermality can then be reached only approximately at
an intermediate time $u=u_t$ defined by
\be
\left. \frac{d}{du} \left< T^f_{uu}(u) \right> \right|_{u=u_t} = 0
\, . \ee Soon after $u=u_t$, however, we loose the exponential
behaviour (\ref{thermal}), and for late times $u \rightarrow
+\infty$ we have a large deviation
\be
\frac{du}{du_{in}} \sim u^2 ( A - B
e^{-Cu} ) \, , \ee
where $A,B$ and $C$ are positive
integration constants depending on $m(v_f)$.\\

In conclusion, during a long time period the radiation is
non-thermal (before and after $u_t$) and the amount of emitted
energy is big enough to allow the information of the initial state
to be released out to future infinity during the evaporation
process. Therefore, back-reaction effects are indeed crucial to
understand the evaporation of a black hole. A more detailed study
requires numerical computations, opening the way to a systematic
analysis to unravel how the details of the incoming matter are
encoded in the outgoing radiation. Finally, we want to stress that,
following the historical analogy, it would also be interesting to
find a theoretical framework capable to reproduce the Hawking flux
proposed here for an evaporating RN black hole.



\end{document}